\documentclass{acm_proc_article-sp}
\usepackage{amssymb}
\usepackage{graphicx}
\usepackage{url}
\usepackage{color}
\usepackage{verbatim}
\usepackage{algorithmic}
\usepackage[ruled, vlined]{algorithm2e}
\newcommand\eg{e.g.}
\newcommand\ie{i.e.}
\newcommand\cf{cf.}

\newcommand\wrt{w.r.t.}
\newcommand\etal{et al.}

\newcommand\stream{\mathcal{D}}
\newcommand\instance{d}

\newcommand\word{w}
\newcommand\classLabel{c}
\newcommand\vocabulary{V}

\newcommand\trainingData{\mathcal{S}}

\newcommand\classifier{\Delta}
\newcommand\ourApproach{ACOSTREAM}

\newcommand\Ji{\textsf{StreamJi}}
\newcommand\TS{\textsf{TwitterSentiment}}

\newdef{definition}{Definition}

\newcommand{\argmax}{\operatornamewithlimits{argmax}}

\newcommand{\eirini}[1]{{\color{black}#1}}

\begin{document}

\title{Incremental Active Opinion Learning Over a Stream of \eirini{Opinionated} Documents}
\numberofauthors{3}
\author{
\alignauthor
Max Zimmermann\\
       \affaddr{Swedish Institute of Computer Science (SICS Swedish ICT)}\\
       \affaddr{E-164 29 Kista, Sweden}\\
       \email{max.zimmermann@sics.se}
\alignauthor
Eirini Ntoutsi\\
       \affaddr{Ludwig-Maximilians University}\\
       \affaddr{Munich 80538, Germany}\\
       \email{ntoutsi@dbs.ifi.lmu.de}
\alignauthor
Myra Spiliopoulou\\
       \affaddr{Otto-von-Guericke University}\\
       \affaddr{Magdeburg 39106, Germany}\\
       \email{myra@iti.cs.uni-magdeburg.de}
}

\permission{This paper was presented at the Fourth International Workshop on Issues of Sentiment Discovery and Opinion Mining (WISDOM 2015), held in conjunction with KDD'15 in Sydney on 10 August 2015. Copyright of this work is with the authors.}

\maketitle          

\begin{abstract}
Applications that learn from opinionated documents\eirini{, like tweets or product reviews,} face two challenges. First, the opinionated documents constitute an evolving stream, where both the \eirini{authors}'s attitude and the vocabulary itself may change. Second, labels of documents are scarce and labels of words are unreliable, because the sentiment of a word depends on the (unknown) context in the author's mind. 
\eirini{Most of the research on mining over opinionated streams focuses on the first aspect of the problem, whereas for the second a continuous supply of labels from the stream is assumed. Such an assumption though is utopian as the stream is infinite and the labeling cost is prohibitive.}
\eirini{To this end, w}e investigate the potential of active stream learning algorithms \eirini{that ask for labels on demand.}
Our proposed \ourApproach\footnote{Source code is available in R at: \url{https://www.dropbox.com/s/y2ptl486f4rvohx/acostream_src.zip?dl=0}} \eirini{approach  works with limited labels: it} uses an initial seed of labeled documents, occasionally requests additional labels \emph{for documents} from the human expert and incrementally adapts to the underlying stream while exploiting the \eirini{available labeled} documents. 
\eirini{In its core, \ourApproach~consists of a MNB classifier coupled with ``sampling" strategies for requesting class labels for new unlabeled documents.} 
In the experiments, we evaluate the classifier performance 
over time 
by varying: (a) the class distribution of the opinionated stream, while assuming that the set of the words in the vocabulary is fixed but their polarities may change with the class distribution; and (b) the number of unknown words arriving at each moment, while the class polarity may also change.\footnote{Datasets are available at: \url{https://www.dropbox.com/s/gcpcyazp7fqentb/streams_acostream.zip?dl=0} }
Our results show that active learning on a stream of opinionated documents, \eirini{delivers good performance while requiring a small selection of labels.}
\end{abstract}


\section{Introduction}
\label{sec:introduction}
New communication media promote sharing social content conveniently, \eg~opinions, ideas, thoughts etc., with everyone connected to
the \emph{WWW}. Blogs, social networks and microblogging are the common services to pose experiences \cite{Cambria:2013}.
Peoples contributions to such services ordered by time of their \eirini{publication constitute a \emph{stream of opinions}}.

An opinion is represented by a document that conveys sentiment; some of its words have a polarity, but these word polarities do not necessarily determine the polarity of the document. 
On the other hand, a word appears in many opinionated documents, and the polarity of these documents gives an indication on whether this word is used to describe positive or negative sentiment. 
Moreover, polarity learning on a stream of documents is driven by scarcity of labeled data, since up to date labeled reviews or tweets are not available -- it is impractical to expect that a human expert inspects and labels arriving reviews or tweets on sentiment, especially in an infinite data stream scenario \cite{Masud:2011}.
In this study, we investigate how the active acquisition of labels on \emph{document} polarity can contribute to learning and adapting upon an ongoing stream of documents.

According to Mohri \cite{Mohri:2012}, the goal of active learning is to achieve a performance comparable to the standard supervised learning scenario, but with fewer labeled examples. Model inference and adaption over streams lends itself to active learning, since the acquisition of fresh labels for all documents of an ongoing fast stream is impracticable. However, learning polarity on streams is subject to two challenges. First, the vocabulary evolves, as new words show up, and as the positive/negative connotation of some words changes. Second, the document polarity model evolves, in the conventional sense of concept drift -- the likelihood of one polarity class becomes higher than before. Most conventional polarity stream mining algorithms, including active learning variants address drift of the document polarity model but assume that the vocabulary is fixed and known in advance \cite{LourencoJr.:SIGIR2014, Guerra:KDD2011}.
\eirini{In this work, we propose an active stream learning approach for evolving feature spaces. 
In the core of our approach, there is a Multinomial Naive Bayes (MNB) classifier, which allows for an easy maintenance of class and word-class statistics over time. } 

In our earlier work~\cite{ZimNtoSpi:NeuroComp14, WagnerPKDD15}, we proposed polarity stream learning algorithms that adapt to an evolving vocabulary in the stream. 
However, in~\cite{WagnerPKDD15} we assume that fresh document labels are made available at each moment, while in~\cite{ZimNtoSpi:NeuroComp14} we assume solely an initial seed of labeled documents and then we adapt the model in a semi-supervised way. On the contrary, in this study, we propose an active stream learning algorithm \eirini{which requests document labels on demands based on the need for adapting the classifier to the underlying stream. }

This work is organized as follows.
Related work is discussed in
Section~\ref{sec:related}. The basic concepts of \ourApproach, the incrementally updating process and the sampling strategies for document label acquisition are presented in Section~\ref{sec:method}. 
Experimental results are shown in Section~\ref{sec:experiments}. Conclusions and open issues are discussed in Section~\ref{sec:conclusions}.

\section{Related Work}
\label{sec:related}
Active learning is a prominent choice when dealing with problems where labeled data are expensive to obtain, \eg~polarity classification or computational biology
applications.
There exist various active learning approaches, provided in recent surveys such as \cite{FuZL13,Settles10activelearning}.
They differ in their heuristics to select instances for which the true label is requested. Garnett \etal~\cite{Garnett2012} use the most likely
or the most pessimistic posterior $P(c|\instance)$ made by a current model. In contrast Krempl \etal~\cite{KremplDS14} and Ho \etal~\cite{Ho:2008} weight the posteriors by their likelihood resp. use hypotheses testing to include the reliability of the posterior when selecting the next instance.
All these approaches follow the same framework: they select the next instance and relearn the classifier with the new instance.
Relearning is expensive in terms of runtime when dealing with large streams as we do.
Our approach works incrementally, thus it does not require relearning rather is expands the current model with new instances.

In context-sensitive learning, it is assumed that the label of a word depends on the context it is used in. Methods that trace recurring concepts \cite{Gama:2014KIS, Lazarescu05} and those that monitor context change \cite{MatKou:SIGMOD10, GuEtAl:WIC11} can trace the association of a word to a label, but only for a limited number of existing contexts, respectively recurring concepts, and for a fixed vocabulary. Therefore, we concentrate on learning with an evolving vocabulary without making assumptions about concept recurrence or context switching.

Zliobaite \etal~\cite{ZliobaiteEtAl:ECMLPKDD2011} propose two sampling strategies which are flexible towards a growing collection as well as considering concept change. The latter is covered while allowing the learner to select also samples which are not close to the decision boundary, \ie~for which the classifier is very certain, so that the classifier will not miss concept change. 
Boy \etal~\cite{Boiy09} test uncertainty and relevance \footnote{Relevance sampling regards the labeling of those examples which are most likely to be class members \cite{Lewis:1994}.} sampling with different classifiers. It
is used to acquire more examples from a class which is scarce. Their results expose that Multinomial Naive Bayes (MNB) classifier performs best for both sampling techniques on polarity classification. We also use MNB as classifier.

Yerva et al.~\cite{Yerva:IJSA2012} propose an active stream learning based classifier for classifying tweets into relevant or irrelevant for
a given company. Their idea is to built a company profile of positive and negative words and test the tweet against the profile to
decide on its class. 
The profile is maintained online over the stream; initially a small set of words is included but the seed set is
expanded by also including words that co-occur often in the stream with words in the seed set.
We also expand in a word-basis, however our approaches are broader rather than topic specific.

Recently Kranjc et al.~\cite{Kranjc2014} present an active learning framework for selecting the most suitable tweets \wrt~an initial
trained classification model. They use as a Support Vector Machine and re-build the model as soon as new suitable tweets are selected. 
They select suitable tweets based on uncertainty and random sampling.
Similarly \cite{Smailovic2014} contribute an active learning approach distinguishing opinionated
(positive and negative) from non-opinionated (neutral) tweets in finance twitter data streams. Based on an SVM classifier, Smailovic et al.
determine a query strategy for active learning, combining advantages from uncertainty and random sampling.

We skip a discussion on the most recent polarity classification algorithms such as Socher et al. \cite{socher2013recursive} as the contribution of our work is towards active learning strategies for polarity classification rather than pure polarity classification.

\section{Active Opinion Stream Learning}
\label{sec:method}
We observe a growing collection $\stream$ of documents that constitute a stream, which we monitor 
at distinct timepoints $t_0,t_1,\ldots,t_i,\ldots$. 
\eirini{Documents arrive at each $t_i$}. 
A document $\instance \in \stream$ is represented by the \emph{bag-of-words} model, \ie~$\instance = { \word_1,\word_2, \cdots, \word_n}$.
We further assume an \emph{initial labeled seed set} $\trainingData$ of documents: for each $\instance \in \trainingData$, an expert has assigned a polarity label $\classLabel \in C$ ($C$ is the set of possible labels, \eg,~{positive, negative}). We borrow the notation of \emph{initial seed set} from our previous work proposed in \cite{ZimmermannEtAl:SAC14}.
As the stream progresses the concept of words might change, \ie~a word which is used to express positive polarity might change its 
contextual relation so that it is used to expressing negative thoughts. Moreover, new words - previously unknown words - might appear as 
peoples' vocabulary to express their positive or negative
opinion evolves over time. 
The mining goal is to assess the polarity label of incoming documents while considering concept change and new words in the stream.

\subsection{\ourApproach~Overview}
\label{sec:OverviewOurApproach}
An overview of our approach is depicted in Algorithm \ref{algo:acostream}.
Briefly, it works as follows:
The seed set $\trainingData$ is used to initially train a classifier $\classifier(\trainingData)$ upon the true labels of $\trainingData$; \eirini{the document labels are propagated to their component words; this way the vocabulary $\vocabulary$ (line 2) is derived. The vocabulary consists of the words observed in $\trainingData$ and their distribution in the positive, negative class. Note that these counts are adequate to approximate the class-conditional word probabilities and the class probabilities in MNB.   
We employ the classifier to predict the label for each arriving new document $d$ from the stream (line 4). 
Depending on the active learning sampling strategy (\cf~Section~\ref{sec:activeLearning}), we might request the true label $c$ for $d$ by an expert (line 7).
If this is the case, we update the related word-class counts and class counts in the model, for all words appearing in $d$ and the true label $c$ of $d$ (lines 8-10). If we encounter some new word, i.e., not in the current vocabulary, we expand the vocabulary accordingly and start monitoring their occurrences in the different classes (lines 10-12).  
Moreover we update the documents-class counts and the seed set while adding documents to $\trainingData$ (lines 13-14).

Note that the classifier's predictions are always made on the current (updated) seed set $\trainingData$. That is, the classifier is a \emph{lazy learner}. Moreover, the seed set consists always of true-labeled documents, i.e., labeling was done by an expert. This implies, that the classifier is always trained upon true labeled (and therefore, reliable) instances.}

\begin{algorithm}[htb!]
\SetAlFnt{\small\sf}
\LinesNumbered
\DontPrintSemicolon
\LinesNumbered

\SetKwData{B}{batch}
\SetKwData{T}{t}
\SetKwData{MNB}{$\Delta$}
\SetKwData{V}{vocabularly}
\SetKwData{setV}{setVocabulary}
\SetKwData{cR}{currentReview}
\SetKwInOut{Input}{Input}
\Input{initial seed $\trainingData$, stream $\stream$}

$\classifier$ $\leftarrow$ train initial classifier on seed $\trainingData$; predictedLabels $\leftarrow \emptyset$\; 
$\vocabulary  \leftarrow$ extract all words from $\trainingData$\;

\While{$\stream$} {
$d \leftarrow$ next document from stream\; 
  $p$ $\leftarrow$ predict label for $d$ by $\classifier(\trainingData)$\;
  \If{$d$ is sampled \wrt~$p$} { 
  c $\leftarrow$ request true label for $d$\;
  \tcp{incrementally update word-class counts}
    \For{i=1 \KwTo $|d|$} {
    \tcp{for existing words}
    \lIf{$\word_i \in \vocabulary$} {
      $N_{ic} = N_{ic}$ + 1 
      }
      \tcp{for new words}
    \Else {
      $N_{ic} = 1$\;
      $\vocabulary \leftarrow \vocabulary \cup \word_i$ \tcp{expand vocabulary}
      }
    }
    $N_{c} = N_{c} + 1$ \tcp{update class counts}
    $\trainingData \leftarrow \trainingData \cup d$ \tcp{update seed set}
    }
  }
  \caption{ACOSTREAM}
\label{algo:acostream}
\end{algorithm}

\eirini{We provide more details in the next subsections.}

\subsection{Building and Maintaining a Polarity Classifier Over Time}
\label{sec:polarityClassifier}
Based on the initial seed set $\trainingData$, we propagate the class labels of the documents to their component words $\word_i \in \vocabulary$, where $\vocabulary$ is the set of words derived from the 
documents $\instance \in \trainingData$. We obtain for each word the word-class counts $N_{i\classLabel}$ stating the number of times $\word_i$ has occurred in 
documents with class label $\classLabel$, \ie
$$
N_{ic} = |\{w_i: \exists d \in \trainingData, \word_i \in d \wedge class(d) = c\}|
$$
We further derive the document class counts $N_{\classLabel}$ expressing the number of documents with label $\classLabel$, \ie
$$
N_{c} = |\{d: class(d)=c \}|
$$
Upon the class and word-class counts we compute the empirical class distributions $\hat{P}(\classLabel)$ \wrt~class $\classLabel$ and the 
empirical word-class distributions $\hat{P}(\word_i|\classLabel)$ for each word $\word_i \in \vocabulary$, as described in the following section. We use a "hat" as in $\hat{P}$ to denote empirical estimates hereafter.

Framing the empirical distributions we build a Multinomial Naive Bayes classifier $\classifier$. It is very fast for induction, robust to 
irrelevant attributes, while providing good prediction performance \cite{McCallum:98}. 
We assess the polarity label of an arriving document $\instance$ while employing $\classifier$ on $\instance$:
\[ class(d) = \argmax\limits_{\classLabel \in \{+,-\}} \hat{P}(\classLabel|\instance) \propto \hat{P}(\classLabel) 
\prod\limits^{|\instance|}_{i=1}\hat{P}(w_i|\classLabel)
\]
That is, the class label of a new document $\instance$ is the one maximizing the posterior probabilities
$\hat{P}(\classLabel|\instance), \classLabel \in C$, which depends on the class conditional probabilities of the words in the document and makes the assumption that these words are independent given the class.

The class prior equals to the ratio of documents in $\trainingData$ labeled as $\classLabel$ and the total number of 
labeled documents $|\trainingData|$, \ie
\begin{equation}
\label{equa:classPriors}
 \hat{P}(\classLabel)=|N_\classLabel|/|\trainingData|
\end{equation}

Analogously, the conditional probability of a word $\word_i$ given a class $\classLabel$ equals to the ratio of documents in $\trainingData$ which are labeled as $\classLabel$ and contain the word $\word_i$. 

\begin{equation}
\label{equa:MNBWordClassProb}
\hat{P}(\word_i|\classLabel) = \frac{N_{i\classLabel} +1 } {\sum^{|\vocabulary|}_{j=1}N_{j\classLabel} +|\vocabulary|}
\end{equation}
We apply the Laplace corrector, $1/|\vocabulary|$, to alleviate the zero frequency problem for words that have not been observed under a given class.

\subsection{Actively Selecting Documents to Acquire New Labels}
\label{sec:activeLearning}
As the stream of documents underlies changes \wrt~the empirical word-class distributions $\hat{P}(\word_i,c)$, the empirical class 
distributions $\hat{P}(c)$ and new appearing words, the initial classifier $\classifier(\trainingData)$ trained upon the initial seed 
set $\trainingData$ might become outdated over time. 
The solution is to update the classifier in order to respond to these changes.
To this end, we incorporate new documents into the seed set $\trainingData$ and accommodate new words to the vocabulary.
We further incrementally update word-class counts $N_{ic}$ and document-class counts $N_c$.
However, we only extend $\trainingData$ by documents which are actively sampled, \ie~for which we requested a true label by an expert.  
\eirini{
There are different techniques for actively sampling labels for new documents; we instantiate our approach with two alternative strategies, one based on information gain and  another based on uncertainty, discussed in the following sections.
Our approach, though, can be coupled with different sampling approaches for labeled document acquisition.}

\subsubsection{Sampling by Information Gain}
\label{sec:InformationGain}
We select a new document $d$ for the extension of $\trainingData$ that shows a gain in information with respect to the thus far observed word-class distribution of words $\word_i \in d$ and the distribution after considering the predicted label for $d$. 
The usage of the information gain is motivated by the attribute selection measures used in decision trees 
\cite{Mitchell:97} and our previous work \cite{ZimmermannICDMW}. It is defined as follows:

\begin{definition}
\label{def:confidence}
[Information Gain]
Let $\instance$ be a new document containing words $\word_i \in d$, for which the current classifier $\classifier$ predicts, for instance, the positive polarity label $+$. The \emph{Information Gain} of $\instance$ \wrt~the predicted label relies upon the difference in entropy before and after the addition of the new label $+$.
\begin{equation}
\label{eq:informationGain}
\mathit{IG}(\instance)=\sum\limits_{\word_i \in \instance \wedge \in \vocabulary} H(N_{i+},N_{i-}) - H(N_{i+} +1, N_{i-})
\end{equation}
Here, $H(N_{i+},N_{i-})$ is the entropy of $\word_i$ regarding the two polarity classes $+$ and $-$, which expresses the purity of the 
class distribution based solely on $\word_i$. 
The second term, $H(N_{i+} +1, N_{i-})$, is the entropy of $\word_i$ when considering $\instance$ and its predicted label, $+$ in this example, as part of the seed set.

The entropy of two positive values $a,b \in \mathbb{N}$ is defined as: 
\[
H(a,b) = - \left[\frac{a}{a+b} * log_{2}(\frac{a}{a+b}) + \frac{b}{a+b} * log_{2}(\frac{b}{a+b} )\right]
\]
\end{definition}

Documents that increase the information reflect the current classifier very well and also enhance the classifiers performance while following the thus far observed word-class distributions, \ie~the distributions become more pure and thus the predictions are less random.
A document that shows a gain in the information \wrt~a predicted label $c$ is sampled, \ie~the true label provided by an expert is requested and then utilized to update the classifier.

We update the classifier based on the received true label. Considering the predicted and the true label for $d$, there are two possible scenarios: 
(i) the predicted label matches the true label, \ie~the classifier is enhanced in its decision when being updated with the true label, and (ii) the predicted label is different from the true label.
The latter case occurs if the classifier does not reflect the current concept underlying the stream, \ie~it makes a wrong prediction. The current concept is assumed to be reflected by the true label of the document. 
Therefore, $\classifier$ must be updated with the true label so that the concept of the related word-class distributions can be changed according to the underlying population of the stream. Hence, we do not miss concept change since we update with the true label.

In case of changes in the word-class distribution, the information gain relies on frequent and old words, \ie~words which have appeared in many documents over time, rather than on words that just newly appeared. This is to be preferred as frequent and old words carry more evidence regarding the class.
A toy example shall help to depict this: assuming a word $w$ that occurred thus far in 30 positive documents, further a new document $d$ appears bearing $w$ and the classifier predicts the negative label for $d$, so there is a change in $w$. The entropy difference for $w$ would be $(1/31) * log_{2}(1/31)$, this is a small value so that it is likely that there is still a gain in information if the class distribution of the other words in $d$ are promoted by the negative label; and thus $d$ is selected to update the classifier.
It is easy to see that the entropy difference regarding $w$ is higher if $w$ has appeared less than 30 times before the change occurs. Hence, we trust more frequent and old words when a change occurs.    

It is noted that when computing the information gain we consider only the entropy difference over words $\word_i \in \instance$, instead of all 
words $\word \in \trainingData$, \ie~we do not iterate over all the words.

\subsubsection{Sampling by Uncertainty}
\label{sec:Uncertainty}
As a second sampling strategy for acquiring true document labels, we utilize \emph{uncertainty}. The idea of uncertainty sampling is to ask the expert for labeling an instance for which the current classifier is less certain, \ie~for which the certainty is below some fixed threshold $\alpha$ \cite{Settles10activelearning}.
Since uncertain examples are close to the classifier's decision border, accommodating them makes the predictions of a classifier more distinctive. 
According to our MNB classifier we use the posterior probability estimates $\hat{P}(+|d)$ and $\hat{P}(-|d)$ computed by MNB as measure for certainty. A low posterior probability means that the classifier is less certain. The uncertainty is then defined as: 

\begin{definition}
\label{def:uncertainty}
[Uncertainty]
Let $\instance$ be a new document and $\classifier(\trainingData)$ be the current classifier that computes the posterior probabilities of the two classes (+, -). The predictions of $\classifier(\trainingData)$ are considered as uncertain if: 

\[
\argmax\limits_{c\in \{+,-\}} \hat{P}(c|\instance) \leq \alpha
\]
where $\alpha$ is a value in (0,1).
\end{definition}
The parameter $\alpha$ is selected manually: small values ensure only few documents to be sampled and thus to update the classifier with documents very close to the decision boundary. 
That is, if the threshold is selected too small then the classifier will miss changes.
In contrast, bigger values assume more examples to be sampled. They also allow sampling of documents that are far from the border and which might bear concept change. This also implies, more label requests though.

\section{Experiments}
\label{sec:experiments}
To evaluate \ourApproach,
we experiment with two real world datasets of opinionated documents (product reviews and tweets). The original streams were modified in order to test the performance of \ourApproach~in extreme and less extreme cases. A detailed description of the datasets is given in Section~\ref{sec:dataset}).
We compared our \ourApproach~against several baselines presented in Section~\ref{sec:baselines}. The results of our experiments are presented in Section~\ref{sec:comparingResults}.  

\subsection{Datasets}
\label{sec:dataset}
Stream \Ji~comes from a dataset first introduced by Yu et al.~\cite{Yu:2011} which contains data crawled from cnet.com, viewpoints. com, reevoo.com and gsmarena.com. 
The true labels of the reviews were derived by the authors from star-ratings. The reviews cover mostly products
and their properties such as ``phone'', ``firmware'' and ``price''. We use only reviews describing single product features,
after removing very short reviews containing less than 2 adjectives.
More details on the dataset and our preprocessing are also provided in our previous work~\cite{ZimmermannICDMW}.
The \Ji~dataset contains 11.374 product reviews and a vocabulary of 3.048 different words. 

Stream \TS, first introduced in~\cite{Go2009}, 
was collected by querying the (non-streaming) Twitter API for messages between April 2009 and June 25, 2009.
The stream is very heterogeneous regarding the content. 
The true labels (ground truth) of the tweets were acquired through the Maximum Entropy classifier using emoticons as class labels.
The stream also depicts a very strong concept shift towards its end, as only one of the two classes, the negative ones, is observed at the end of the stream.
The original stream contains 1.600.000 tweets; we focus on the last part of the stream, tweets 1.235.000 - 1.485.000, reflecting concept drift. The selected dataset consists of 250.000 tweets with a vocabulary of 169.853 different words.

In \Ji~we focused only on adjectives and adverbs for sentiment analysis since, according to \cite{Turney:ACL2002,Yu:EMNLP2003}, these words bear the actual opinion of the author; similar observation were shown in~\cite{ZimmermannICDMW}. Stream \TS~comes with nouns and verbs as stated in \cite{Go2009}. 

\subsubsection{The effect of new appearing words}
\label{sec:newAppearingWords}
In our experiments we show how \ourApproach~performs in a continuously expanding vocabulary $\vocabulary$,i.e., when new words arrive over time from the stream.
To this end, we re-order the original streams so that the number of appearances of words from the initial seed set $\trainingData$ decreases over time whereas the number of new words increases. That is, vocabulary-wise the initial seed set becomes ``outdated" w.r.t. the evolving stream. The ordering was done as in our previous work~\cite{ZimmermannICDMW}.

\begin{figure}[htbp]
\centering
\includegraphics[width=0.45\textwidth]{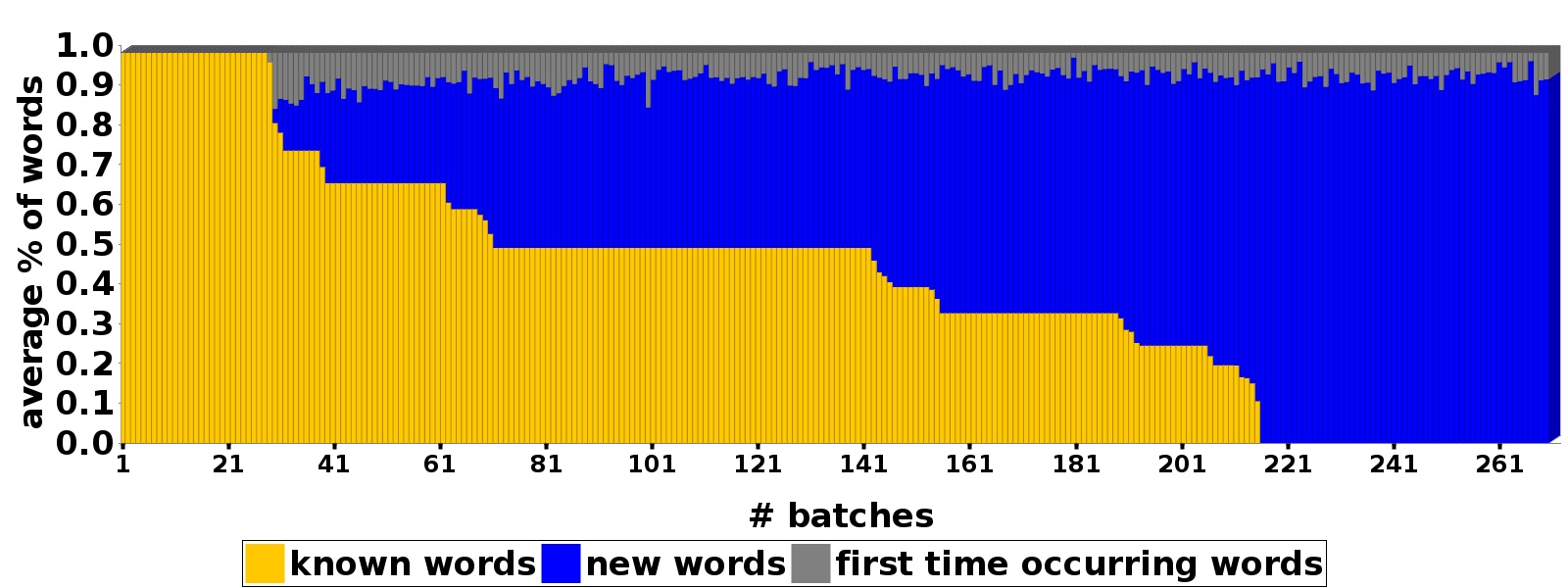}
\includegraphics[width=0.45\textwidth]{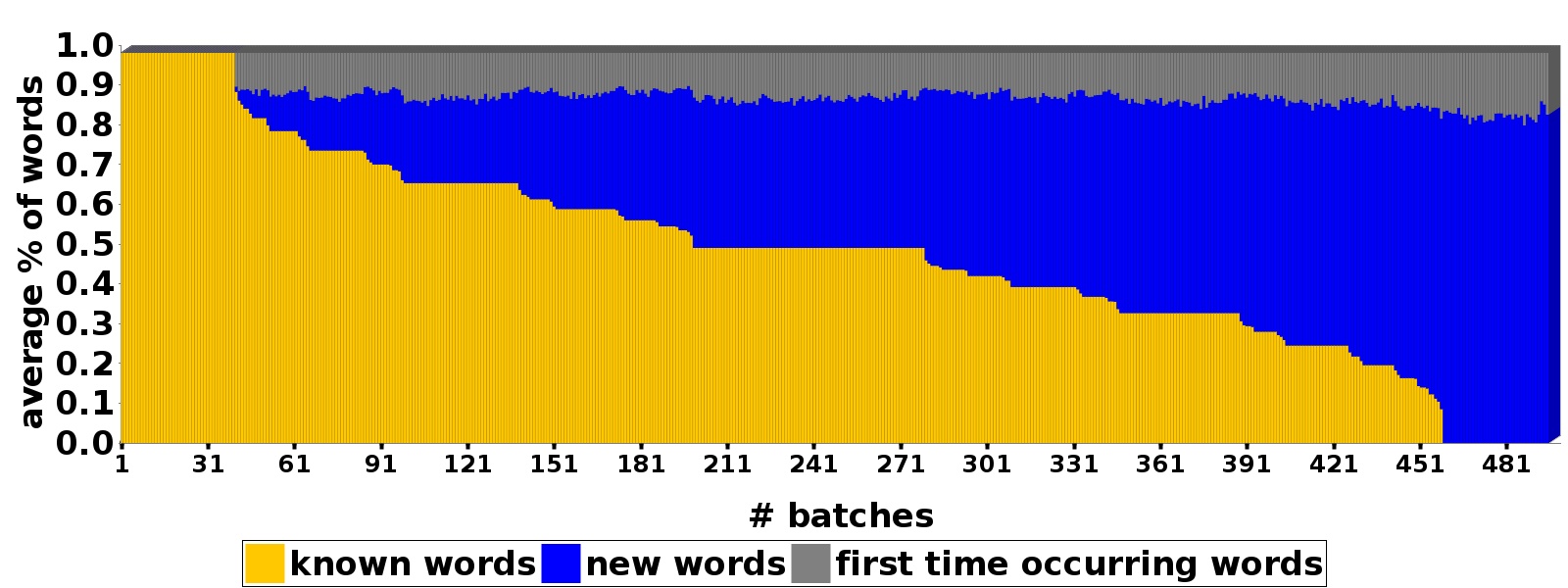}
\caption{\small{Percentage of known, new  and first appearing words over time (avg. per batch) for the re-ordered version of stream \Ji~(top)
$|\trainingData|$=140 resp.\TS~(bottom)
$|\trainingData|$=5.000}}
\label{fig:newWords}
\end{figure}

Based on the ordering procedure, we obtain for each original stream a re-ordered counterpart which begins with documents that contain only words from the initial vocabulary $\vocabulary$ extracted from $\trainingData$; as the stream progresses, the number of new words increases while documents arrive that also contain words $\word \notin \vocabulary$.
In Figure~\ref{fig:newWords} 
we draw the percentage of known and new words per document over time for the re-ordered versions of the streams averaged over batches of size 42 resp. 5000.
In the very beginning all words are known, over time though, the ratio of known words decreases with unknown words dominating the stream.

We distinguish the unknown w.r.t. to the initial seed set words into i) first-time observed new words (in gray) and ii) already monitored new words (in blue).
In the \emph{re-ordered} versions in Figure \ref{fig:newWords}, the number of new words is increasing over time and after some point the stream bears
merely new words; whereas the number of first-time observed words is rather static over time showing a continuously increasing variety of
words. 
The reason for re-ordering is to show how the classifier deals with an expanding vocabulary. 

\begin{figure}[htbp]
\centering
\includegraphics[width=0.49\textwidth]{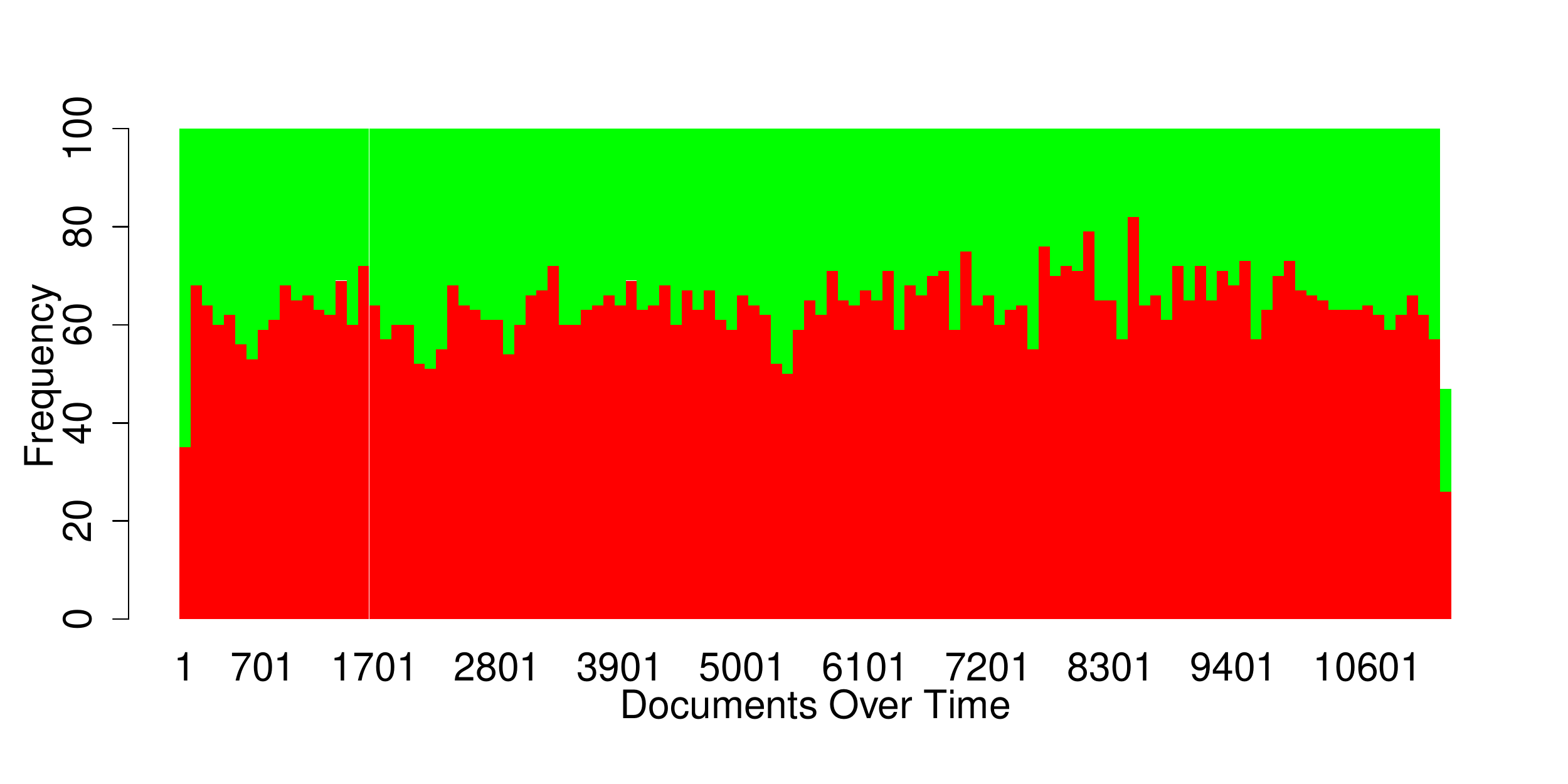}
\includegraphics[width=0.49\textwidth]{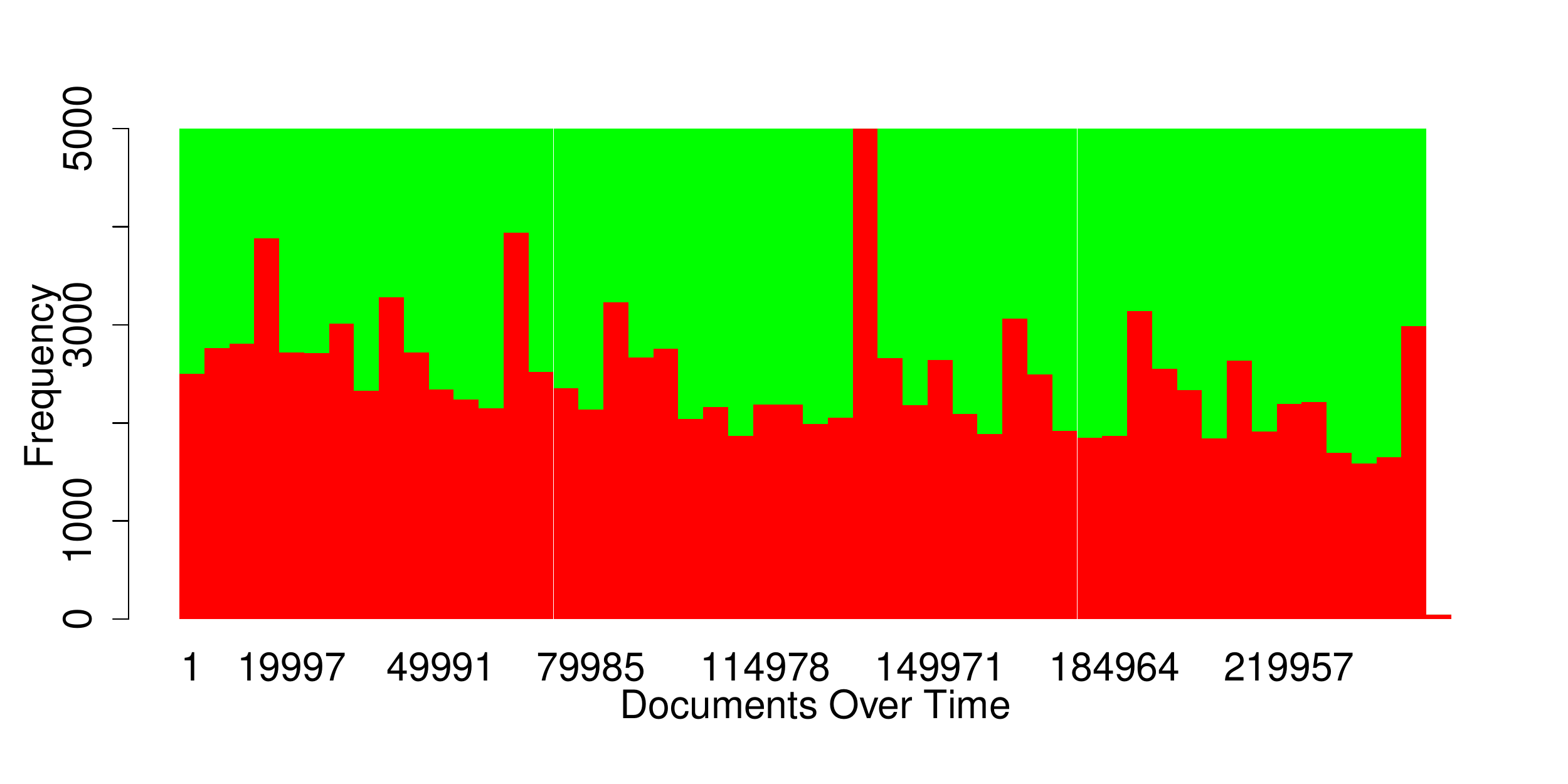}
\caption{\small{Class distribution on \Ji~(top) and \TS~(bottom) accumulated over batches of size 100 resp. 5.000}}
\label{fig:distrDocuments}
\end{figure}

The class distributions of the streams is depicted in Figure \ref{fig:distrDocuments}: \Ji~is slightly skewed towards the negative class over the whole stream while \TS~is uniformly distributed at the beginning whereas, as the stream progresses, the distribution moves more towards the positive class.  

\begin{figure}[htbp]
\centering
\includegraphics[width=0.49\textwidth]{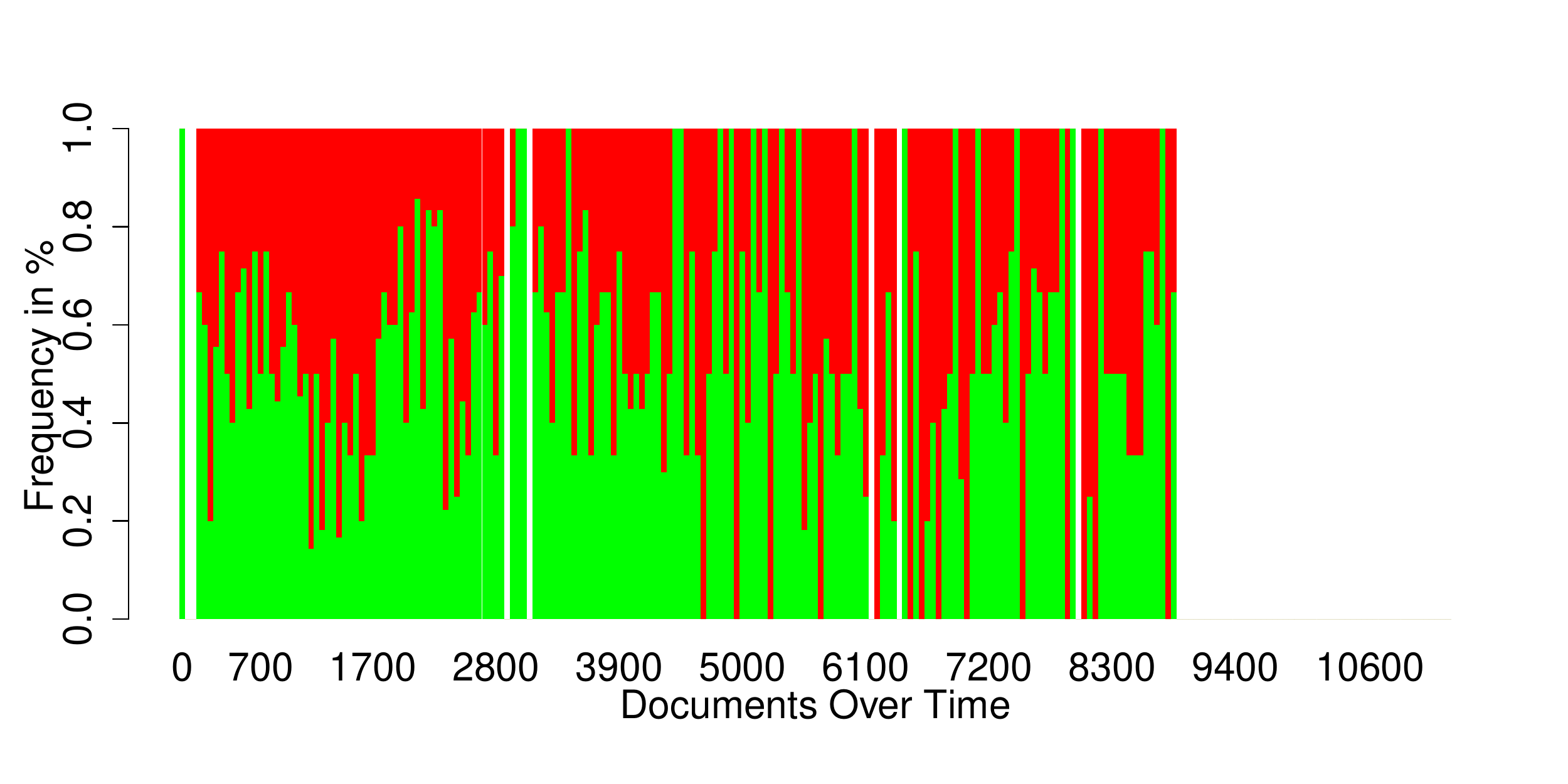}
\includegraphics[width=0.49\textwidth]{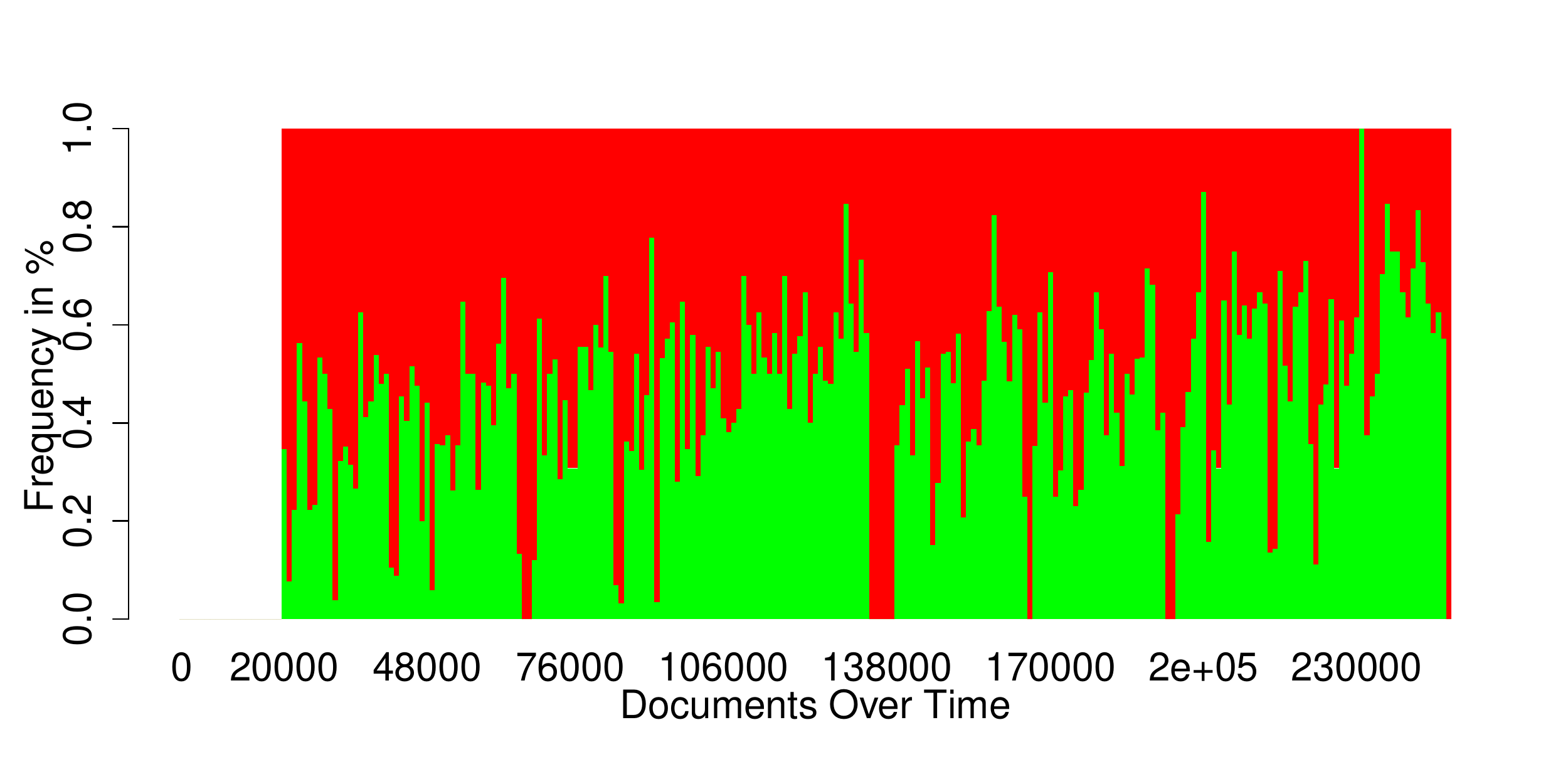}
\caption{\small{Word-class distribution of the words ``best" on \Ji~(top) and ``tomorrow" on \TS~(bottom) accumulated over batches of size 50 resp. 1.000 and depicted as frequency in percentage}}
\label{fig:distrWords}
\end{figure}

The obtained re-orderings of the original streams bear also changes in the polarity of words, \ie~the word-class distributions changes over time. Figure \ref{fig:distrWords} depicts the word distribution of the words ``best'' on \Ji~and ``tomorrow'' on \TS~as accumulated ratio of documents with positive(green) resp. negative label(red): the distribution of both words change over time, \eg~for word ``tomorrow'', the ratio of negative documents alternates heavily as for instance, at document 13.800 only negative documents are shown followed by a majority of positive documents.

\subsubsection{Fixed Vocabulary}
\label{sec:fixedSetOfWords}
The scenario where new words appear over time is an extreme one; though, it is a rather realistic one in polarity learning over streams.
To apply our approach on a less extreme scenario, we run experiments on streams showing up \emph{NO} new words over time, \ie~the seed contains all words of the stream. 

\begin{figure}[htbp]
\centering
\includegraphics[width=0.49\textwidth]{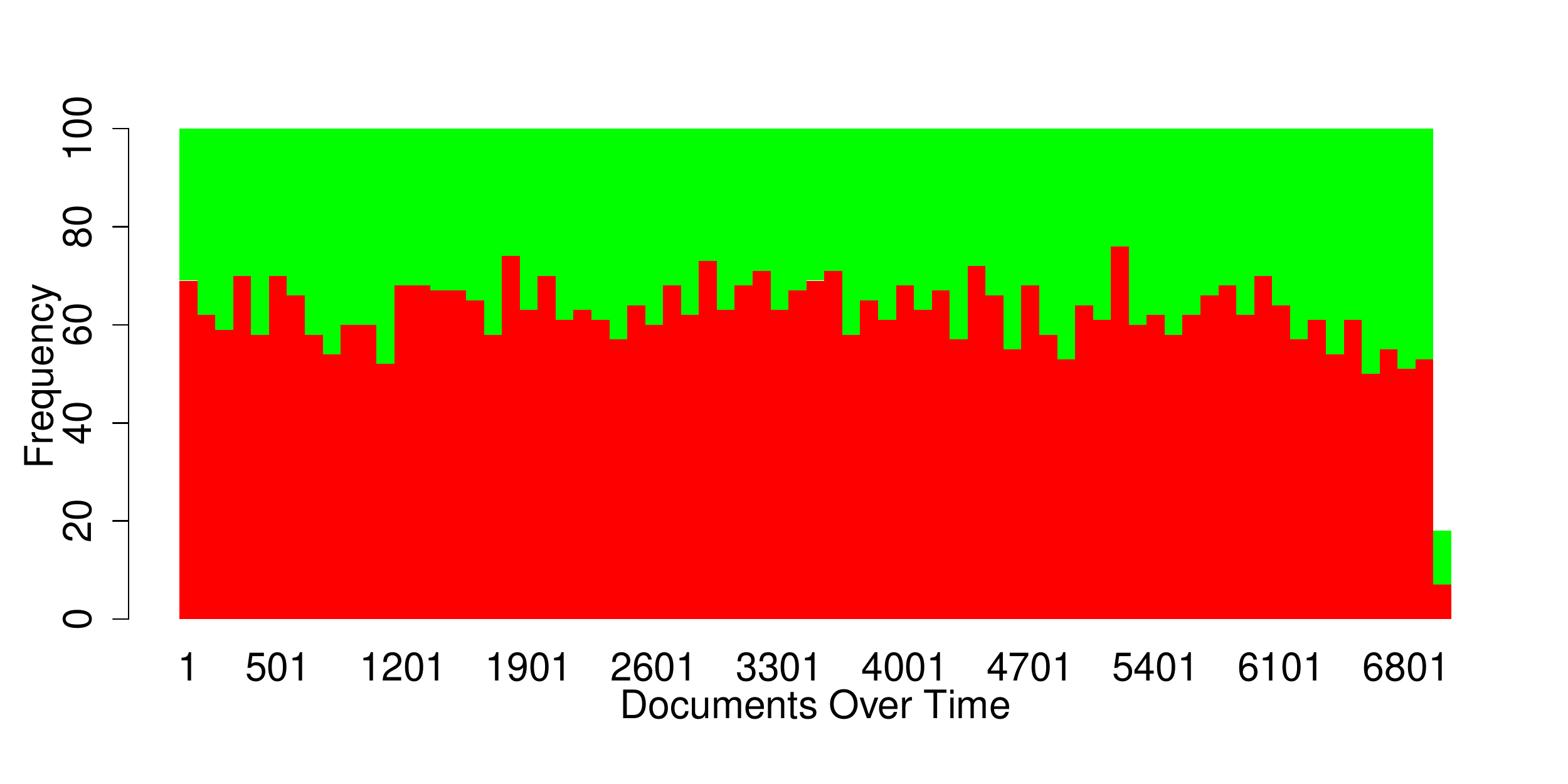}
\includegraphics[width=0.49\textwidth]{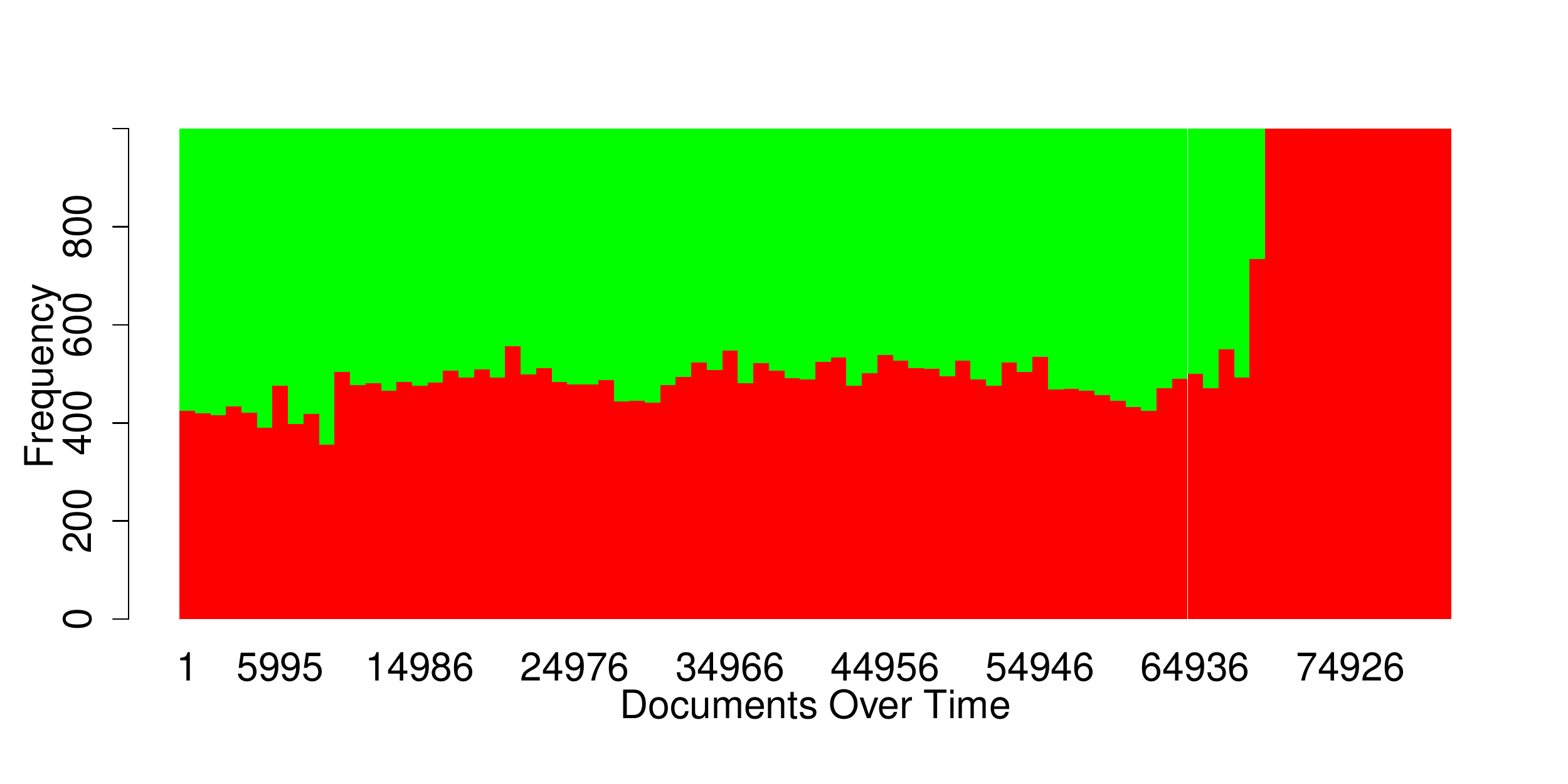}
\caption{\small{Class distribution of streams \Ji~and \TS~showing no new words over time \wrt~seeds with sizes of 1.000 resp. 10.000 documents}}
\label{fig:distrStream}
\end{figure}

Therefore, we reduced the original streams, keeping the \emph{original order} though, to documents that contain only words which are part of the initial vocabulary $\vocabulary$ extracted from the seed.
We acquired the shortened stream while selecting a relatively large seed $\trainingData$ (1.000 for \Ji~and 10.000 for \TS). Based on $\trainingData$, we extracted the vocabulary $\vocabulary$, and as the stream progresses we considered the documents $\instance$ that contain only words $\word \in \vocabulary$. 

The class distribution of the constituted streams is depicted in Figure \ref{fig:distrStream}. We aggregated the number of positive and negative documents over batches of size 100 and 1000 for \Ji~resp. \TS. The resulting versions of the stream are smaller than the original version: \Ji~contains 7.018 documents and 759 words while \TS~covers 81.480 tweets and 14.785 words. 

Similar to the re-ordered versions of the streams, described in \cf~Section~\ref{sec:newAppearingWords}, the shortened stream bears concept change of the words. We skip detailed figures on specific words though as they mostly conform with the word distributions depicted in Figure \ref{fig:distrWords}.

\subsection{Learning methods and quality measures}
\label{sec:baselines}
Below we outline the approaches we used to compare against \ourApproach.
They all use Naive Bayes as classifier but differ on which documents they use for adaptation.

\begin{itemize}
 \item \textbf{IncrementalMNB:}
 The classifier is updated gradually with each incoming instance based on the true labels of the instances. It assumes 100\% availability of true labels. This approach serves as an upper baseline.
 \item \textbf{StaticMNB:}
The classifier is not updated over time, rather is is trained once upon the initial seed set and remains static over the whole stream. This approach  serves as a lower baseline.  
 \item \textbf{Random:}
 The random sampling strategy labels the incoming instances at random instead of deciding actively on the relevance of the label.
 For every incoming instance the true label is requested with a probability $B$, where $B$ is the budget \cite{ZliobaiteEtAl:ECMLPKDD2011}. We switch the budget in our experiments among 0,3 and 0,6, e.g., 30\% of the documents from the stream are asked for the true label.  
\end{itemize}

To evaluate the quality of our classifiers, we use the \emph{kappa statistic}, which normalizes the classifier's accuracy by the accuracy of a chance classifier:
$k = \frac{p_0-p_c}{1-p_c}$ \cite{BifetEtAl:DS2010}.

$p_0$ is the accuracy of a classifier and $p_c$ is the probability of making a correct prediction by a chance classifier
that assigns the same number of examples to each class as the classifier under consideration. 
The kappa varies among -1 and 1: a value $\leq 0$ indicates that the classifier's predictions coincide with, or are worse,
than the predictions of the chance classifier. A value $> 0$ implies that the classifier's predictions overcome these of
a chance classifier. The higher the value, the more often the predictions match with the true labels. Kappa is preferred to accuracy for data streams as it can handle imbalanced class distributions.

\subsection{Performance evaluation}
\label{sec:comparingResults}
In this section, we compare \ourApproach~using \emph{information gain} and \emph{uncertainty} sampling strategies against the \emph{IncrementalMNB}, the \emph{StaticMNB} as well as the \emph{random} sampling based on the performance of kappa over time. As we deal with an evolving stream of documents a fixed budget of true labels cannot be utilized which is normally applied when comparing across different sampling strategies \cite{ZliobaiteEtAl:ECMLPKDD2011}. This would, however, lead to an unfair comparison as the budget would be spent differently among the strategies.
Rather we used different values for the uncertainty threshold $\alpha$ and for random sampling across our experiments yielding to different number of requested labels over the stream. 
We depict the number of requested labels over the stream in percentage of the stream length in Table \ref{tab:requestedDocuments}: \emph{IncrementalMNB} always asks for 100\% of the labels, while \emph{StaticMNB} uses only the true labels of the training set $\trainingData$.
We implemented two experiments: i) we kept the vocabulary fixed over the stream while considering documents that contain only words $\word \in \vocabulary$, \cf~Section~\ref{sec:newAppearingWords}, and ii) we allow the set of words $\vocabulary$ to evolve as including new appearing words.

\begin{table*}[Htbp!]
 \begin{tabular}{|p{4.4cm}||c|c|c|c|c|}
 \hline
  Experiment + Dataset & \ourApproach~(IG) & \ourApproach~(U) & IncrementalMNB & StaticMNB & Random  \\
  \hline \hline
  fixed $\vocabulary$: \Ji &44  & 40 &100 &1 &40\\ \hline
  fixed $\vocabulary$: \TS& 40 & 47 & 100 & 1 & 42\\ \hline
  evolving $\vocabulary$: \Ji & 60 & 59 &100 & 1 & 60 \\ \hline
  evolving $\vocabulary$: \TS & 52 & 88& 100& 2& 31 \\ \hline
 \end{tabular}
\caption{Requested labels per method and experiment: numbers in percentage regarding the length of the stream including the documents to train the classifier, \ie~the size of the seed. \small{(IG=Information Gain), (U=Uncertainty)}}
\label{tab:requestedDocuments}
\end{table*}

In the following we examine the performance of \ourApproach~on the two experiments comparing against the baselines described in Section~\ref{sec:baselines}. 

\subsubsection{Results on the Fixed Vocabulary Stream}
\label{sec:resultsFixedVocabulary}
\begin{figure}[htbp!]
\includegraphics[width=0.49\textwidth]{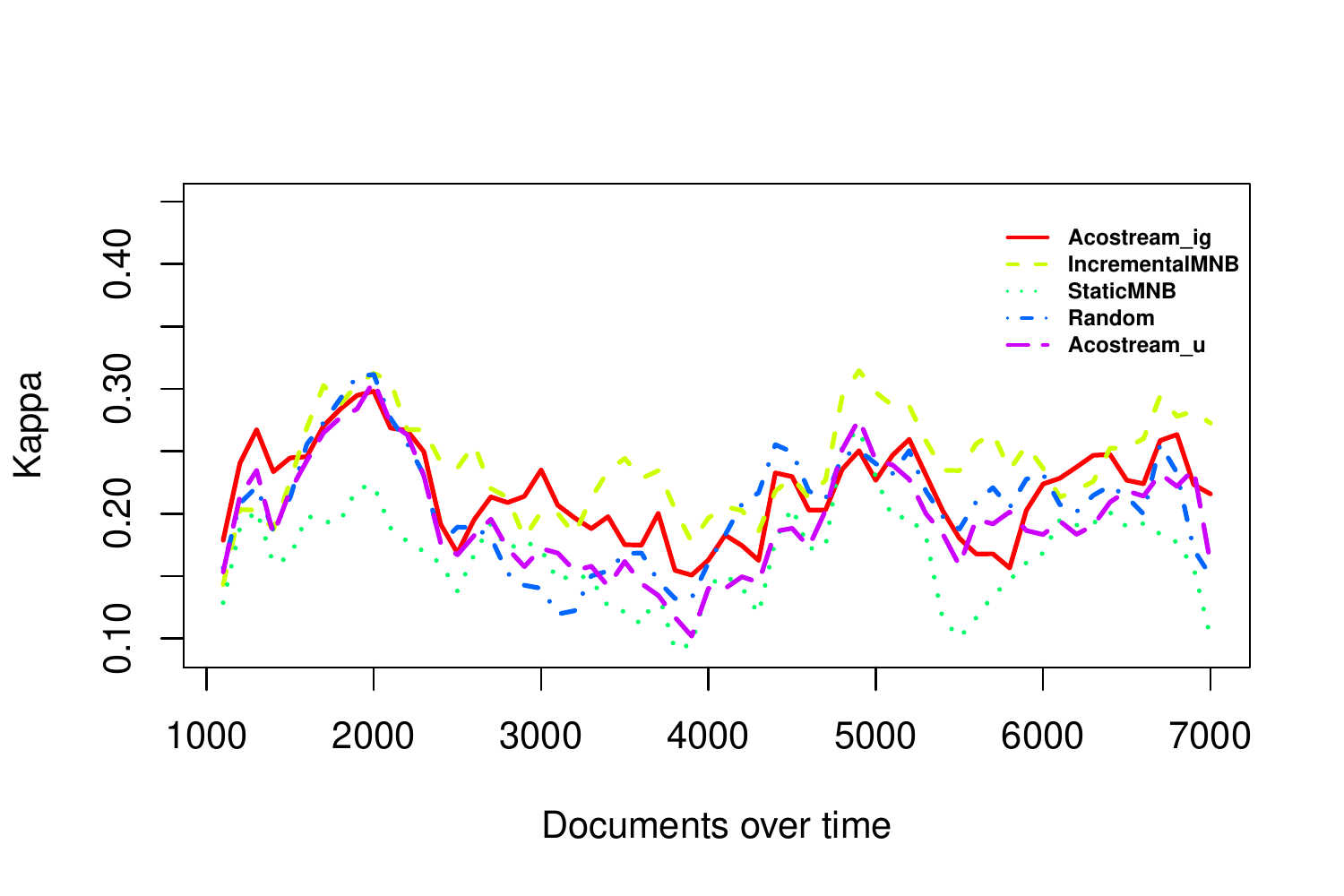}
\includegraphics[width=0.49\textwidth]{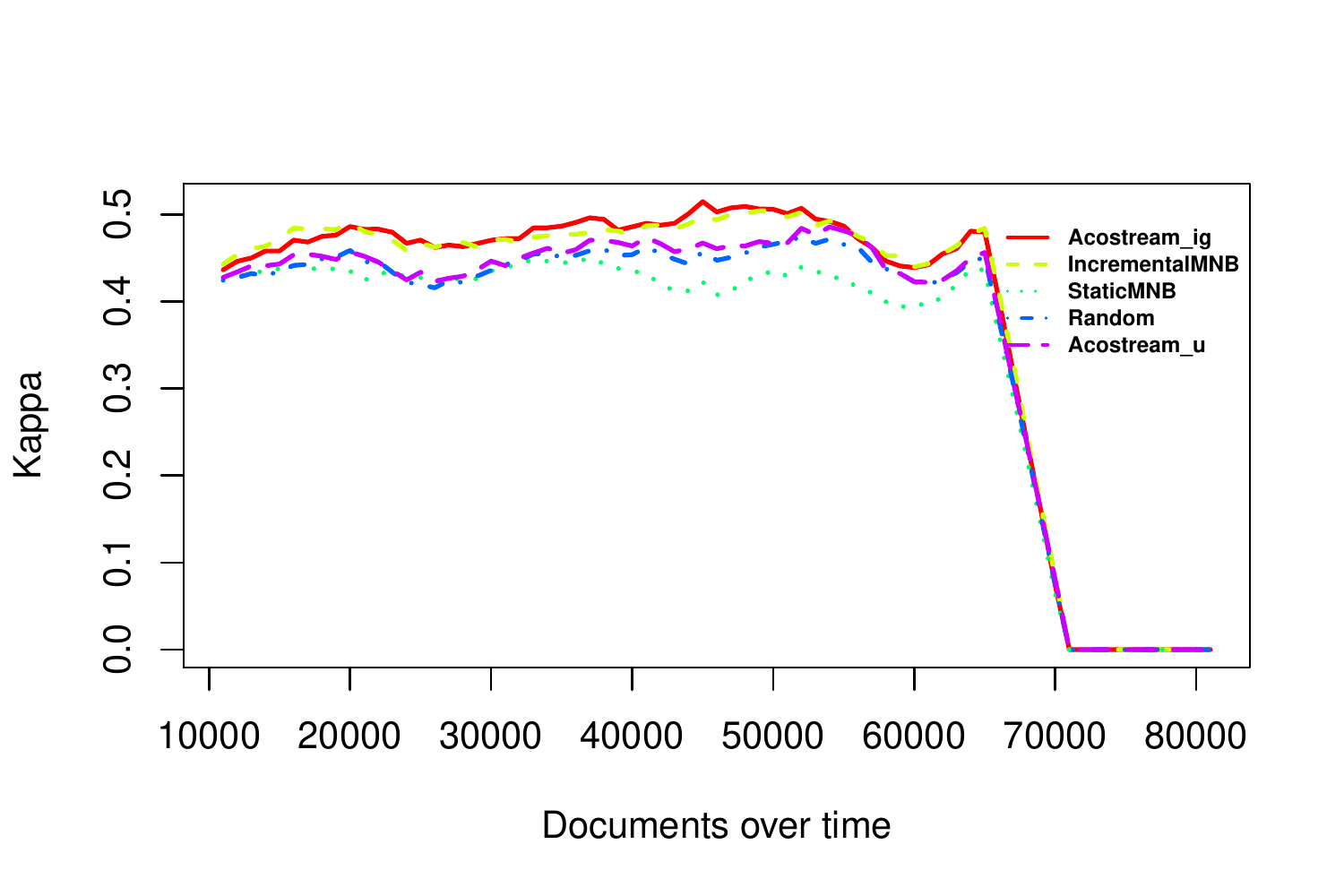}
\caption{Kappa for the three methods to which we compare and \ourApproach~on
stream \Ji~(top) and \TS~(bottom) with a fixed vocabulary}
\label{fig:ExperimentsFixedWords}
\end{figure}

We report on the results carried out from our experiments upon streams with a fixed set of vocabulary and evolving word-class counts, \cf~Section~\ref{sec:fixedSetOfWords}, and while using kappa as evaluation measure. Figure~\ref{fig:ExperimentsFixedWords} depicts the kappa over time for \ourApproach~using information gain (\emph{Acostream\_ig}) and uncertainty (\emph{Acostream\_u}) sampling, \emph{IncrementalMNB}, \emph{staticMNB} and \emph{Random} sampling on the shortened streams \Ji~(upper picture) and \TS~{lower picture}.

\ourApproach~shows a good performance on both streams when applying information gain sampling. The results expose, upon stream \Ji, a kappa that is rather close to the kappa of the upper baseline while utilizing only 44\% of the true labels (\cf~Table~\ref{tab:requestedDocuments}); on \TS~it overcomes the \emph{IncrementalMNB} in most times of the stream using only 40\% of the labels.  
Hence, information gain sampling performs very well requesting only 40\% resp. 44\% of the labels to achieve a comparable or higher kappa than 
\emph{IncrementalMNB} which samples 100\% of the labels.
In contrast, uncertainty sampling, which uses 40\% resp. 47\% of the labels on \Ji~resp. \TS, shows a lower kappa similarly to the results obtained by random sampling.
The reason why kappa drops to 0 at the end of \TS~is because there only negative documents arrive, consequently one cannot be better than a chance classifier.

\subsubsection{Results on continuously expanding vocabulary}
\label{sec:resultsEvolvingVocabulary}
\begin{figure}[htbp!]
\includegraphics[width=0.49\textwidth]{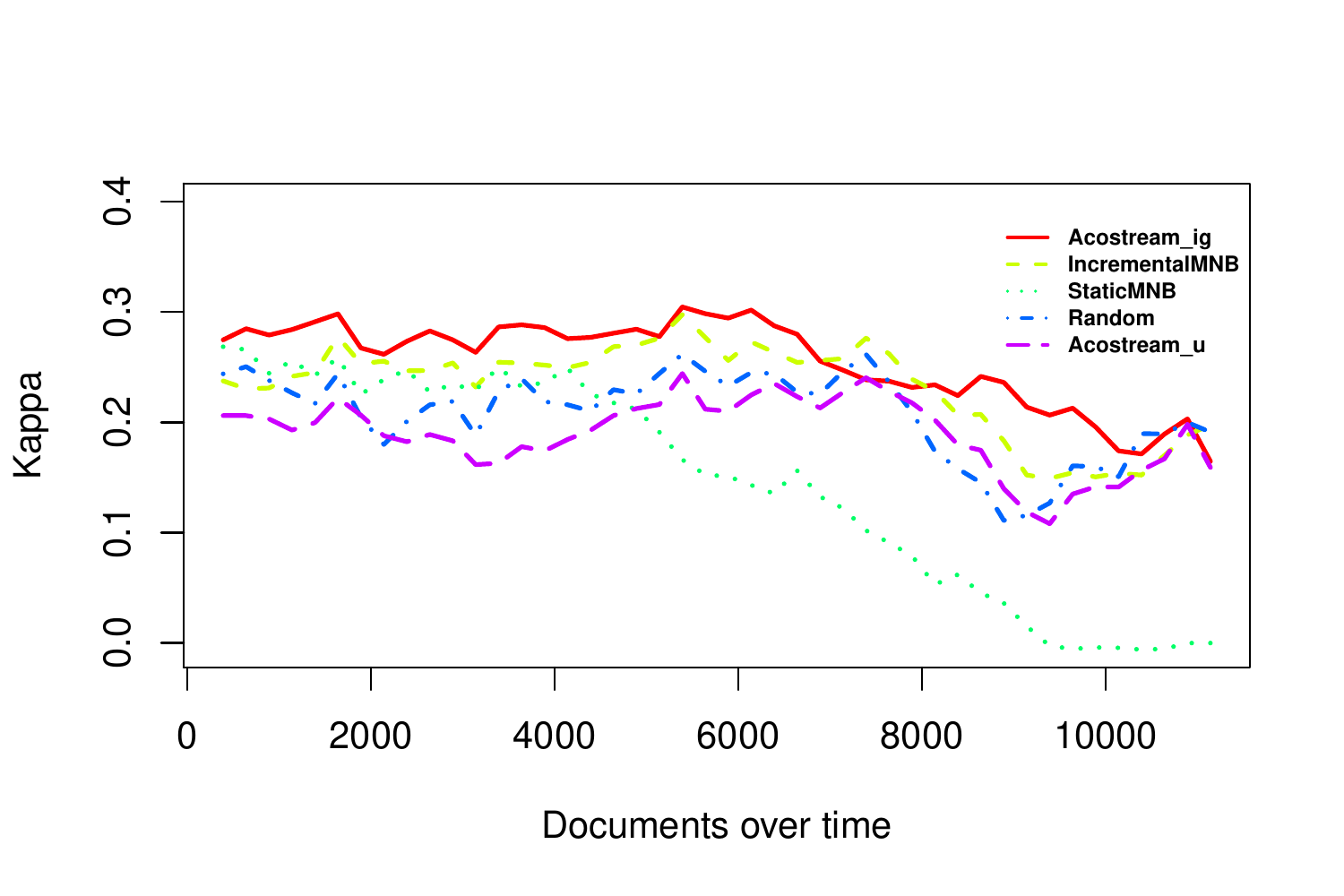}
\includegraphics[width=0.49\textwidth]{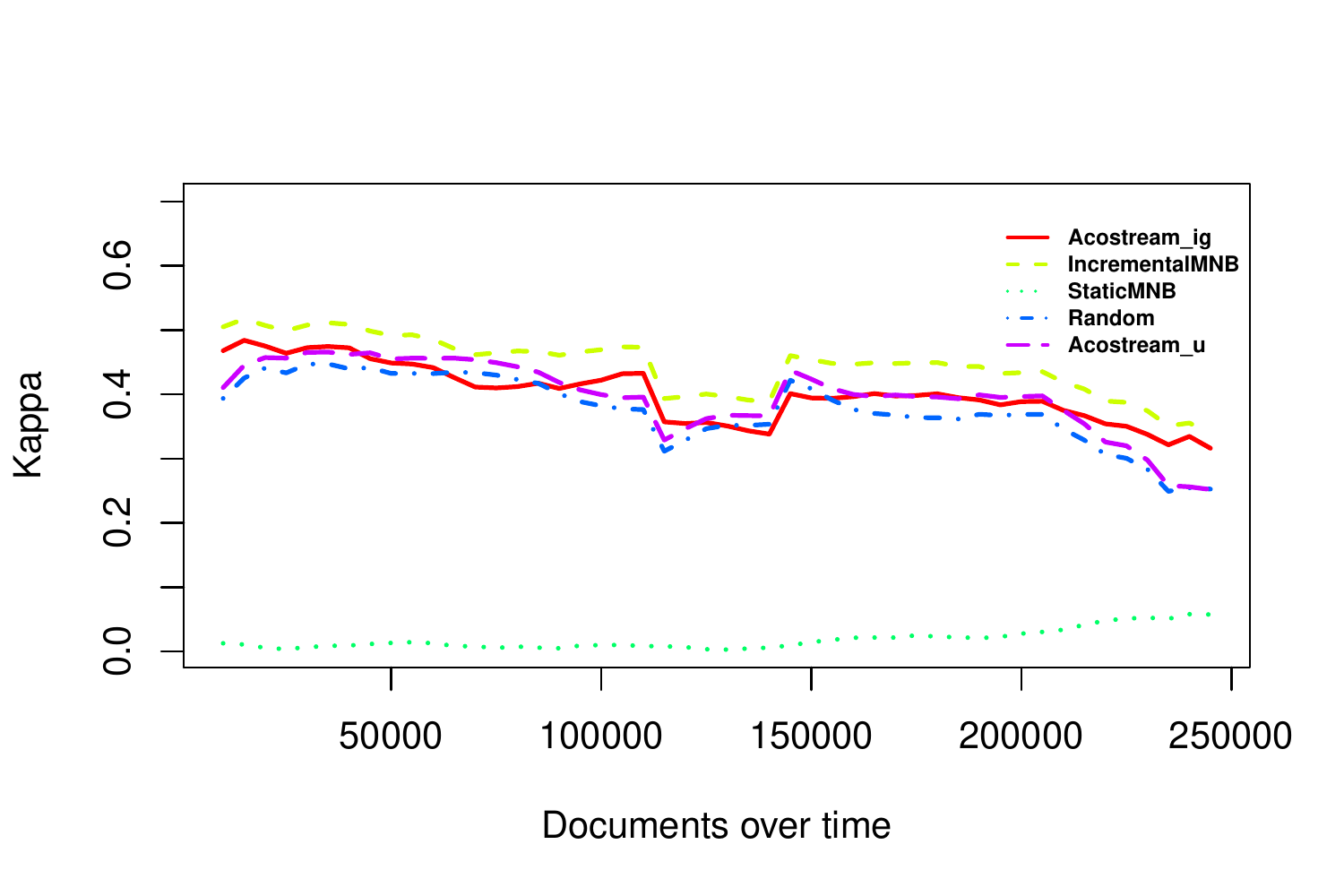}
\caption{Kappa for the three methods to which we compare and \ourApproach~on
stream \Ji~(top) and \TS~(bottom) under an evolving vocabulary}
\label{fig:ExperimentsNewWords}
\end{figure}

We examine how the performance of \ourApproach~is affected by a continuously expanding vocabulary and evolving word-class distributions.   
On stream \Ji~information gain sampling performs very well showing the highest and most robust kappa over time among all approaches to which we compare using only 60\% of the labels, depicted by the picture on top of Figure \ref{fig:ExperimentsNewWords}, 
Uncertainty sampling does not perform well on stream \Ji~showing a lower kappa in comparison to random sampling. On \TS~it performs well but requiring 88\% of the labels to be competitive with \ourApproach~when using information gain sampling that acquires only 52\% of the labels.
The results on stream \TS, depicted by the bottom picture of Figure \ref{fig:ExperimentsNewWords}, reveal that \emph{IncrementalMNB} performs best on large streams with many words (169.853). \ourApproach~(both sampling strategies) follows while showing a similar pattern of the kappa curve but with slightly lower values.

\ourApproach~is not negatively affected by new words and exposes a stable performance across both streams. Also, the curves show a pattern similar to the one obtained from the \emph{IncrementalMNB} that adapts with all documents of the stream and thus considers all changes of the word distributions. That is, \ourApproach, in particular when information gain is used, adapts well to the underlying change in the population of the stream.  

\subsubsection{Effect of the uncertainty threshold $\alpha$}
\label{sec:resultsUncertainty}
\begin{figure}[htbp]
\includegraphics[width=0.49\textwidth]{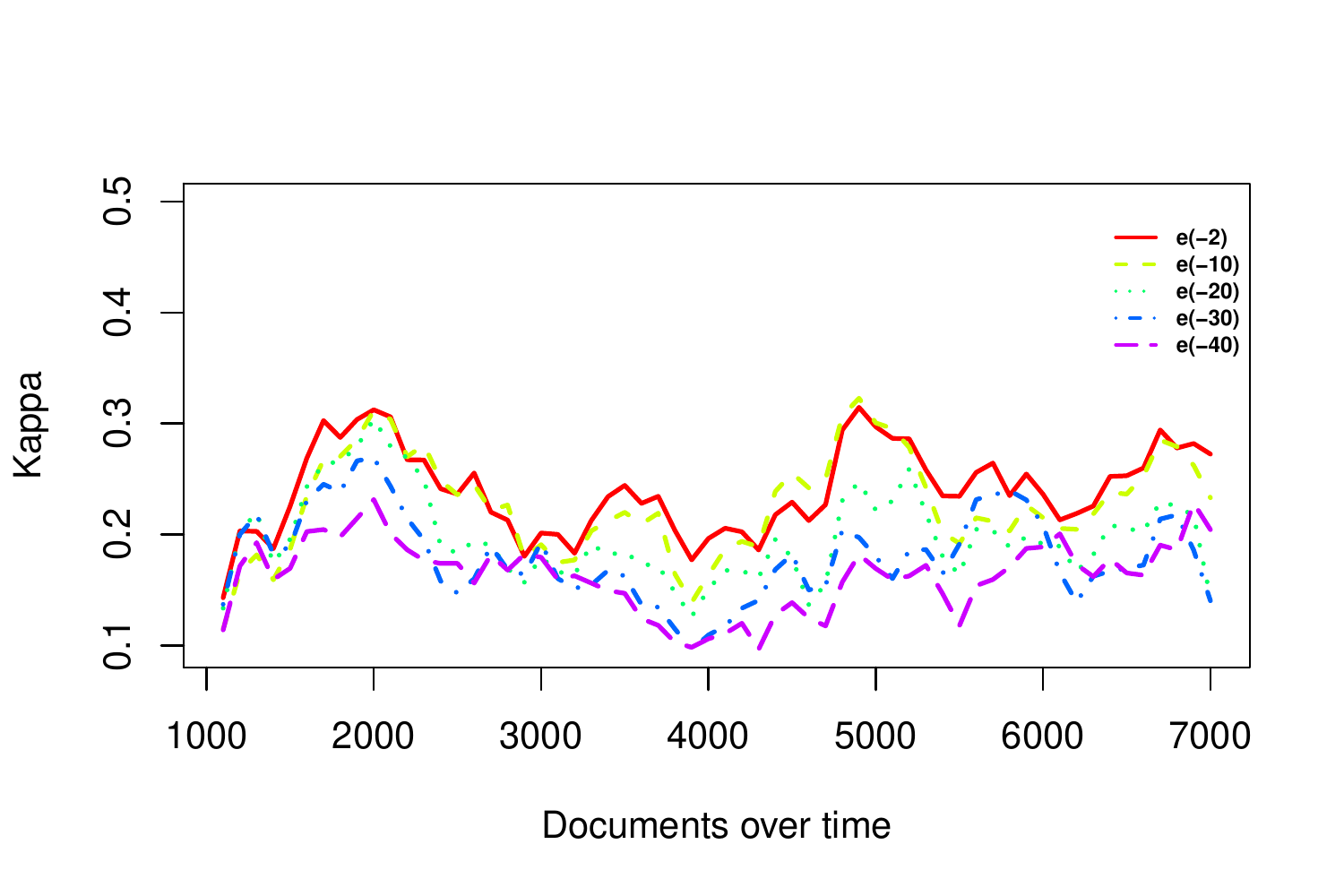}
\includegraphics[width=0.49\textwidth]{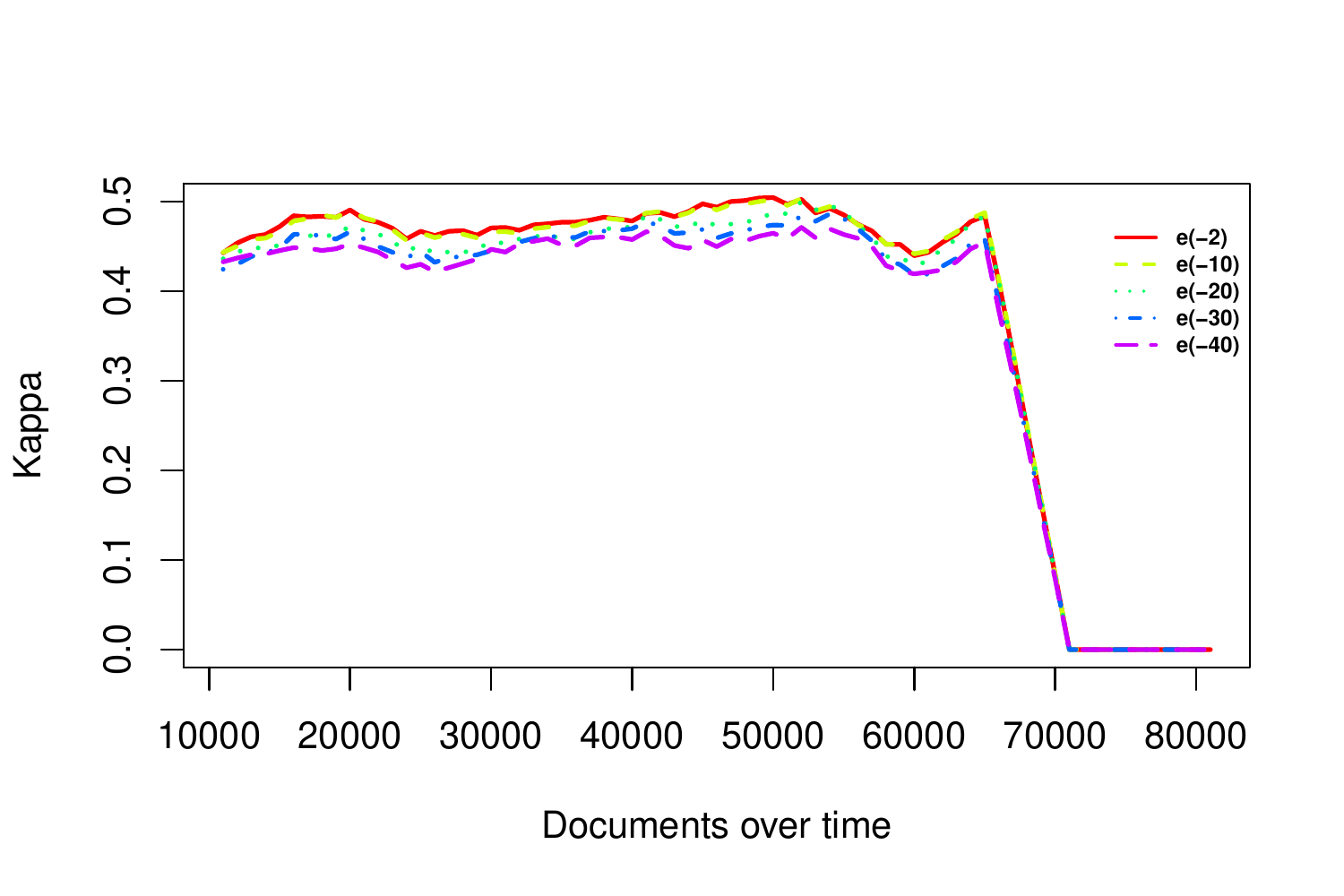}
\caption{Kappa for different settings of $\alpha$ when using uncertainty as sampling strategy for \ourApproach~on
stream \Ji~(top) and \TS~(bottom)}
\label{fig:ExperimentAlpha}
\end{figure}

To show the effect of the uncertainty threshold $\alpha$, \cf~Section \ref{sec:Uncertainty}, we varied values of $\alpha$ on stream \Ji~and \TS~when the vocabulary has a fixed size. Figure~\ref{fig:ExperimentAlpha} depicts kappa over time on \Ji~(upper picture) and \TS~(lower picture) for different settings of $\alpha$ when uncertainty is used as sampling strategy upon \ourApproach. We varied among five values: e(-2), e(-10), e(-20), e(-30) and e(-40), where \emph{e()} is the exponential function. Note that the posteriors become rather small as dealing with a sparse feature space. Thus we had to set small values for $\alpha$ in order to cause difference in the consumption of labels.   

The results on both streams show and increasing performance while taking larger values for $\alpha$ into account. This is not surprising, as with increasing $\alpha$ also the number of considered samples grows which intuitively leads to a better performance.  
The gap in performance among values for $\alpha$ is huge on stream \Ji~where 100\%, 87\%, 33\%, 19\%, and 15\% percent of labels are requested; while on \TS~ 100\%, 92\%, 74\%, 55\% and 40\% of the documents are sampled, leading to smaller gaps between the curves.

\section{Conclusion}
\label{sec:conclusions}
Polarity learning on an evolving stream is a challenging task as the stream is subject to concept changes; existing words might change sentiment over time due to e.g., different context, but also new words might occur to express opinions.
Another challenge for a stream polarity learner is the scarcity of the class labels, assuming manual labeling of the (infinite) stream is unrealistic. Responding to these challenges requires adaptation of the model to the underlying stream population based on only a few labeled examples.	

In this work, we proposed our active stream learning framework \ourApproach~for incrementally updating a polarity learner based on actively acquired document labels. We instantiate our framework with two sampling strategies, \emph{information gain} and \emph{uncertainty}. 
We compare our method to 
a traditional active learning approach (random sampling), an incremental approach that requires all arriving document labels and a non-adaptive method. 
Our results show that actively asking for labels, pays off as the performance of the classifier is quite good while the label consumption remains low.
Comparing the two sampling approach, \emph{information gain}-based sampling shows good performance on all datasets w.r.t. the number of required labels, the accuracy of predictions and adaptation to concept change. The \emph{uncertainty}-based sampling on the contrary shows a poor performance. 

Our ongoing work involves more elaborated techniques on propagating document labels to words, considering that not all words contribute the same to the polarity of a document. 
Furthermore, we want to diminish independence between new documents when deciding to sample them. This will allow us to sample in a wider prospect detecting change early and address emerging scenarios comprehensively.
Moreover, we plan to instantiate \ourApproach~with different classifiers (except for the currently employed MNB) and different sampling strategies for active learning.

\section{Acknowledgment}
The work of Max Zimmermann was carried out during the tenure
of an ERCIM ``Alain Bensoussan'' Fellowship Programme.

 \bibliographystyle{abbrv}
\bibliography{references}
\balancecolumns
\end{document}